\RequirePackage[2020-02-02]{latexrelease}

\documentclass{article}
\usepackage{spconf,amsmath,graphicx}
\usepackage{amssymb}
\usepackage{mathtools}
\usepackage{multirow}
\usepackage[table, dvipsnames]{xcolor}

\usepackage[printwatermark=true]{xwatermark}
\newwatermark*[page=1,color=gray!60,fontfamily=cmr,fontseries=m,scale=0.4,xpos=0,ypos=120, width=2\paperwidth]{This paper has been accepted for publication at the \\ IEEE International Conference on Acoustics, Speech and Signal Processing (ICASSP), Singapore, 2022}
\newwatermark[page=1,color=gray!60,fontfamily=cmr,fontseries=m,scale=0.3,xpos=0,ypos=-125, width=2.7\paperwidth,align=left]{\textcopyright\ 2022 IEEE. Personal use of this material is permitted. Permission from IEEE must be obtained for all other uses, in any current or future media, including reprinting/republishing this material for advertising or promotional purposes, creating new collective works, for resale or redistribution to servers or lists, or reuse of any copyrighted component of this work in other works.}

\definecolor{myblue}{rgb}{0.82, 0.95, 0.93}

\title{T-NGA: Temporal Network Grafting Algorithm for Learning to Process Spiking Audio Sensor Events}
\name{Shu Wang\textsuperscript{\dag}\qquad Yuhuang Hu\textsuperscript{\dag}\thanks{\dag Equal contribution.}\qquad Shih-Chii Liu\thanks{This work was partially funded by the SNSF grant 
CRSII5\_177255 and the Swiss National Competence Center in Robotics (NCCR Robotics). We thank Sheng Zhou for help with the Python software cochlea model.}}

\address{Institute of Neuroinformatics, University of Z\"urich and ETH Z\"urich,
Z\"urich, Switzerland}

\begin{document}
\ninept
\maketitle
\begin{abstract}
Spiking silicon cochlea sensors encode sound as an asynchronous stream of spikes from different frequency channels. The lack of labeled training datasets for spiking cochleas makes it difficult to train deep neural networks on the outputs of these sensors. This work proposes a self-supervised method called Temporal Network Grafting Algorithm (T-NGA), which grafts a recurrent network pretrained on spectrogram features so that the network works with the cochlea event features. T-NGA training requires only temporally aligned audio spectrograms and event features. Our experiments show that the accuracy of the grafted network  was similar to the accuracy of a supervised network trained from scratch on a speech recognition task using events from a software spiking cochlea model. Despite the circuit non-idealities of the spiking silicon cochlea, the grafted network accuracy on the silicon cochlea spike recordings was only about 5\% lower than the supervised network accuracy using the N-TIDIGITS18 dataset. T-NGA can train networks to process spiking audio sensor events in the absence of large labeled spike datasets.
\end{abstract}
\section{Introduction}
With the increasing use of speech command interfaces on mobile devices, finding lightweight networks for edge audio tasks, \emph{e.g.}, Voice Activity Detection (VAD), Speaker Verification (SV), and Keyword Spotting (KWS), has become a fast growing research area~\cite{rohit2015}.
Recent methods to produce low-memory footprint or low-compute networks include the use of delta activation update~\cite{gaoDigit2019} and Neural Architectural Search~\cite{zhang2021autokws} methods.

\begin{figure}[ht]
    \centering
    \includegraphics[width=\linewidth]{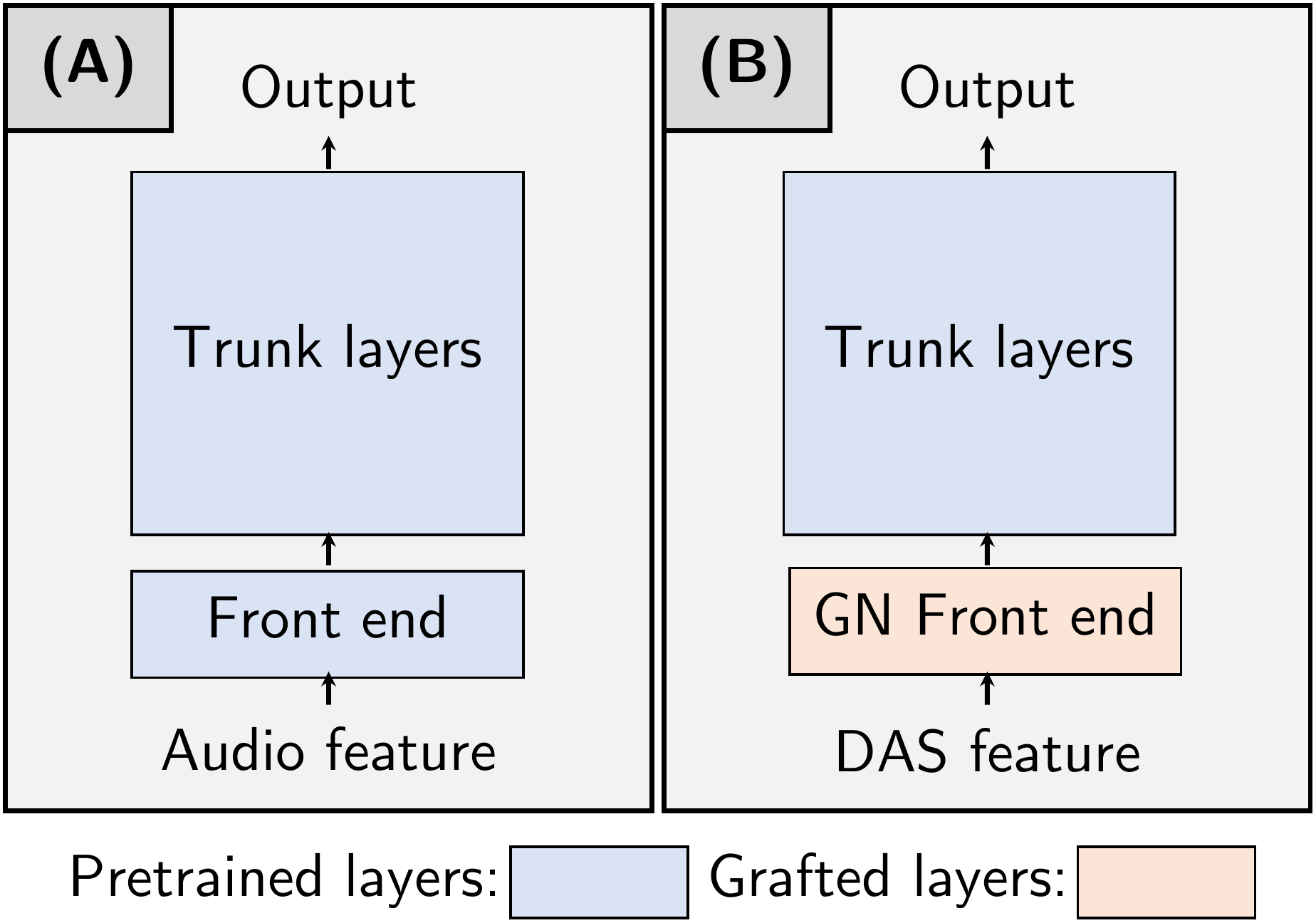}
    \caption{Temporal Network Grafting Algorithm (T-NGA). ``GN'' stands for grafted network.}
    \label{fig:tnga:front:page}
\end{figure}

The inputs to these networks are typically spectrogram features extracted from the audio samples, \emph{e.g.}, Mel-Frequency Cepstral Coefficients (MFCCs) and Log-Mel spectrograms. On the other hand, spiking silicon cochlea sensors, which implement a simple model of the biological cochlea, encode audio as an asynchronous stream of output spikes from a set of frequency channels. The spiking Dynamic Audio Sensor (DAS) incorporates filter circuits that model the frequency selectivity of the basilar membrane, the rectification of inner hair cells, and the spike generation of the auditory nerve fibers~\cite{das:Liu:2014, das:Yang:2016}.
The DAS output events have been used to drive low-latency localization solutions together with the separation of the spike streams produced by competing talkers~\cite{Anumula2018}.
These studies show that it is possible to simultaneously localize multiple speakers, to separate their spike streams, and to estimate the speech envelope of an individual speaker from the separated spike streams. The DAS output has been used for edge audio tasks such as SV~\cite{ceolini2019event} and digit recognition~\cite{gaoDigit2019} where the low-power and low-latency properties of an edge audio device are needed. Other methods use front-end software models that produce audio events, which are then processed by a spiking neural network
~\cite{wu2020deep,yilmaz2020deep,martinelli2020spiking}. Custom edge Application-Specific Integrated Circuit (ASIC) designs that include the VAD chip by~\cite{yangbnn2018} use a spiking cochlea front-end together with a multilayer binary perceptron, and demonstrate a low power consumption of 1\,$\mu$W.

Training deep neural networks for audio processing often requires large labeled datasets such as LibriSpeech which contains over 960 hours of transcribed audio~\cite{libre:speech:Panayotov:2015}. There is a lack of large labeled spike datasets for event-driven sensors, thereby preventing the more extensive use of deep networks for processing spiking cochlea output events. To mitigate the expensive costs for labeling spiking audio datasets, we propose a self-supervised Temporal Network Grafting Algorithm (T-NGA), which requires only temporally aligned audio waveforms and spike recordings for training. In other words, T-NGA training does not require labeled DAS recordings.
In this work, we show how T-NGA can exploit a pretrained network that processes audio spectrogram features to produce a \emph{grafted network} (GN) that processes event features via Transfer Learning. T-NGA extends the Network Grafting Algorithm (NGA)~\cite{nga:Hu:2020} which was originally proposed for vision modalities to temporal audio sequences. 

We studied and validated T-NGA using a Recurrent Neural Network (RNN) on a simple Automatic Speech Recognition (ASR) task (Section~\ref{sec:results}), first using spike recordings generated from an ideal software spiking cochlea model and then from the DAS recordings in the N-TIDIGITS18 dataset~\cite{event:feature:Anumula:2018}. We investigated whether the accuracy of the GN obtained from T-NGA training could achieve the accuracy of a network trained from scratch using the labeled spiking cochlea recordings; and how the silicon circuit non-idealities and recording non-idealities affect the accuracy of GN.
The contributions of this work can be summarized as follows:
\begin{itemize}
    \item We propose a self-supervised training method, T-NGA, which avoids the need of expensive labeling when training on novel spiking cochlea events for an acoustic task such as ASR.
    \item T-NGA, to our knowledge, is the first algorithm that utilizes temporally aligned audio spectrogram and event features for training a network that uses spiking cochlea events for an ASR task.
\end{itemize}

\section{Methods}
\label{sec:methods}

\subsection{Temporal Network Grafting Algorithm (T-NGA)}\label{subsec:tnga}

Fig.~\ref{fig:tnga:front:page} illustrates the working principle of T-NGA. An RNN $\mathcal{F}(\cdot)$ is pretrained on spectrogram features extracted from the audio waveform samples. The pretrained network $\mathcal{F}(\cdot)$ consists of two parts: a front end $\mathcal{F}_{\text{front}}$ and trunk layers $\mathcal{F}_{\text{trunk}}$ (Fig.~\ref{fig:tnga:front:page}~(A)). The weights of the pretrained network are frozen during the T-NGA training. After training, a newly-trained grafted front end $\mathcal{G}_{\text{front}}(\cdot)$ together with the pretrained trunk layers $\mathcal{F}_{\text{trunk}}$ form a \emph{grafted network} (GN) that processes event features computed from the spiking cochlea events (Fig.~\ref{fig:tnga:front:page}~(B)).

Let $\{(\mathbf{a}_{i}, t_{i}^{a})\}_{i=1}^{T^{a}}$ and $\{(\mathbf{s}_{t}, t_{i}^{s})\}_{i=1}^{T^{s}}$ be the temporally aligned spectrogram and event features where $t_{i}^{a}$ and $t_{i}^{s}$ are timestamps of $\mathbf{a}_{i}$ and $\mathbf{s}_{i}$ respectively. Note that $T^{a}$ may not be equal to $T^{s}$ depending on the choice of feature configuration (see Section~\ref{subsec:audio:feature}).

$\mathcal{F}(\cdot)$ processes the spectrogram features $\{(\mathbf{a}_{i}, t_{i}^{a})\}_{i=1}^{T^{a}}$ and returns the prediction $\{y_{l}\}_{l=1}^{L^{a}}$:
\begin{align}
    \{\mathbf{h}_{i}\} = \mathcal{F}_{\text{front}}(\{\mathbf{a}_{i}\}),\qquad \{y_{l}\} = \mathcal{F}_{\text{trunk}}(\{\mathbf{h}_{i}\})
\end{align}
where $\{(\mathbf{h}_{i}, t_{i}^{a})\}_{i=1}^{T^{a}}$ are the pretrained states, \emph{i.e.}, hidden states of $\mathcal{F}_{\text{front}}$. Similarly, GN processes event features $\{(\mathbf{s}_{t}, t_{i}^{s})\}_{i=1}^{T^{s}}$ and outputs prediction $\{y_{l}^{s}\}_{l=1}^{L^{s}}$:
\begin{equation}
    \{\mathbf{g}_{i}\} = \mathcal{G}_{\text{front}}(\{\mathbf{s}_{i}\}),\qquad \{y_{l}^{s}\} = \mathcal{F}_{\text{trunk}}(\{\mathbf{g}_{i}\})
\end{equation}
where $\{(\mathbf{g}_{i}, t_{i}^{s})\}_{i=1}^{T^{s}}$ are the grafted states, \emph{i.e.}, hidden states of $\mathcal{G}_{\text{front}}$. The dimension of $\mathbf{g}_{i}$ is the same as the dimension of $\mathbf{h}_{i}$.

T-NGA training maximizes the similarity between $\{\mathbf{h}_{i}\}$ and $\{\mathbf{g}_{i}\}$. When $T^{a}\neq T^{s}$, we select $T$ features from both $\{\mathbf{h}_{i}\}$ and $\{\mathbf{g}_{i}\}$ where $T\leq\min(T^{a}, T^{s})$ and timestamps of the corresponding states $t_{i}^{a}$ and $t_{i}^{s}$ are the closest in time. T-NGA training optimizes the following loss function:
\begin{align}
    \mathcal{L}&=\text{NCS}(\{\mathbf{h}_{i}\}, \{\mathbf{g}_{i}\})+\text{MAE}(\{\mathbf{h}_{i}\}, \{\mathbf{g}_{i}\}) \\
    &=1+\frac{1}{T}\sum_{i=1}^{T}\frac{\mathbf{h}_{i}\cdot\mathbf{g}_{i}}{\|\mathbf{h}_{i}\|\cdot\|\mathbf{g_{i}\|}}+\frac{1}{T}\sum_{i=1}^{T}|\mathbf{h}_{i}-\mathbf{g}_{i}|\label{eqn:tnga:loss}
\end{align}
where $\text{NCS}(\cdot, \cdot)$ and $\text{MAE}(\cdot, \cdot)$ compute the negative cosine similarity and the mean absolute error between aligned states, respectively. When the training is completed, the grafted front end $\mathcal{G}_{\text{front}}$ encodes the event features into a set of grafted states that can be directly used by the pretrained trunk layers $\mathcal{F}_{\text{trunk}}$.

\begin{figure}[ht]
    \centering
    \includegraphics[width=\linewidth]{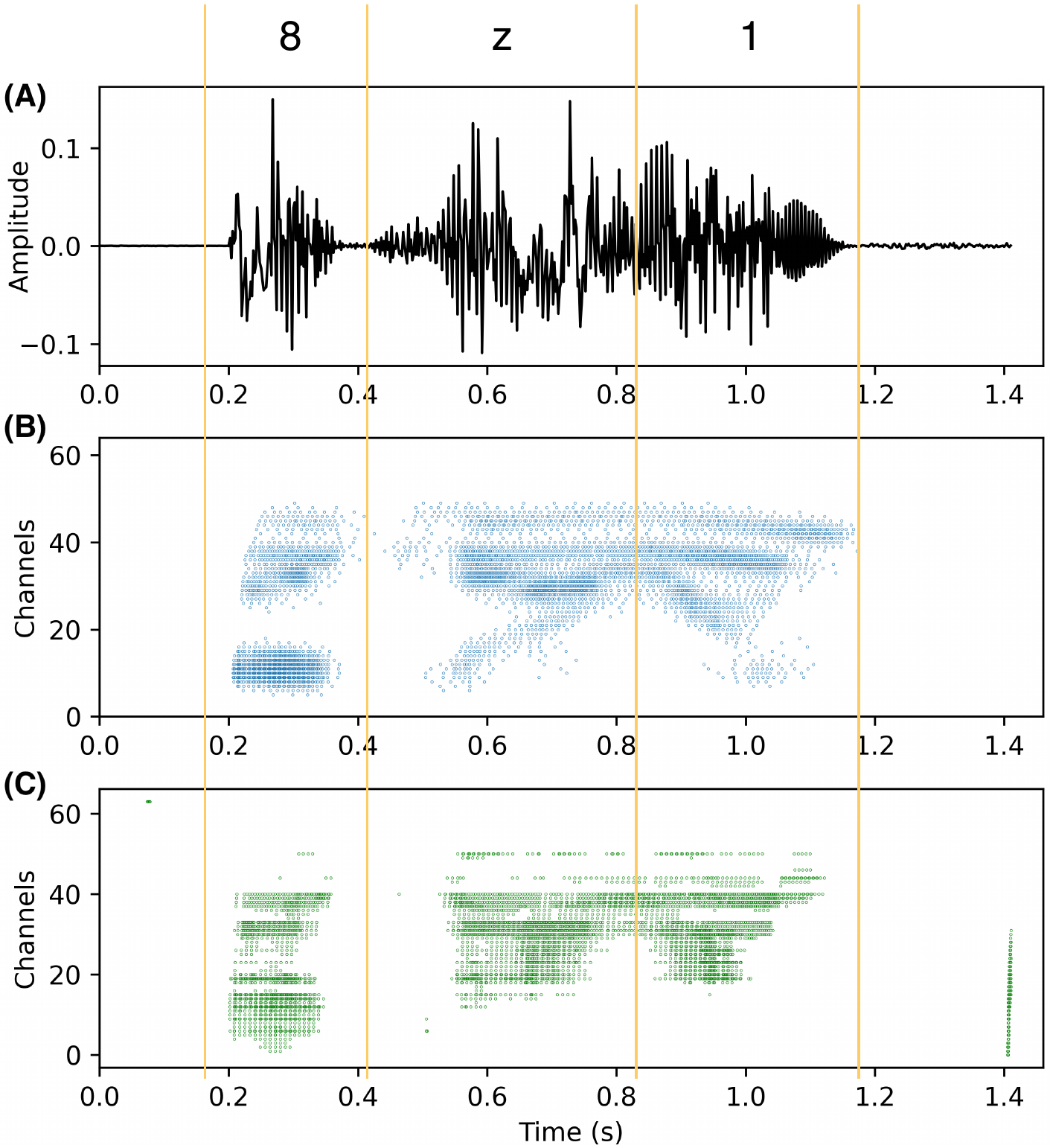}
    \caption{Time waveform \textbf{(A)} of a sample ``8z1'' from the TIDIGITS dataset. Also shown are spikes generated using an ideal software cochlea model \textbf{(B)}, and real DAS spikes \textbf{(C)} from the 64 cochlea frequency channels in response to the audio sample.}
    \label{fig:audio:events}
\end{figure}

\subsection{Spiking Cochlea}\label{subsec:das}

The Dynamic Audio Sensor (DAS) has two independent 64-stage cascaded filter banks driven by two microphone inputs. The frequency selectivity of the cochlea channels ranges roughly from 50\,Hz to 20\,kHz on a log spacing across the channels. The DAS has been captured in several cochlea designs~\cite{das:Yang:2016,das:Liu:2014,aer:Chan:2007}.
The spiking cochlea designed in \cite{aer:Chan:2007} was used to generate the spike recordings of the TIDIGITS dataset to create a spiking cochlea dataset called N-TIDIGITS18. Details of this spiking dataset are given in Sec~\ref{subsec:dataset}.

In this work, we also use a software model of a frequency channel which includes a second-order filter, a half-wave rectifier, and a linear leak integrate-and-fire neuron model used in the design reported in~\cite{aer:Chan:2007}.
The transfer function of the $n$-th filter in a cascaded set of second-order filters is given by
\begin{equation}
    H_n(s)= \prod_{i=1}^{n}  \left(\frac{\tau_i\,s}{\tau_i^2\,s^2 + \tau_i\,s/Q + 1}\right)
\end{equation}
where $\tau_i=\frac{1}{2\cdot\pi\cdot f_i}$, $f_i$ is the center frequency of the $i$-th channel, and $Q$ is the filter quality factor.
The half-wave rectification is described as:
\begin{equation}
    v_r=\max(0, V_n-V_{\text{ref}})
\end{equation}
where $v_r$ is the rectifier output used to drive the neuron, $V_n$ is the filter output of the channel and $V_{\text{ref}}$ is a constant. The parameters of the model are set so that the spike responses of the channels are similar to those of the hardware spiking cochlea.

An example of the spikes generated from the software spiking cochlea model and the spikes recorded with the hardware spiking cochlea in response to the sample ``8z1'' (8Z1A.WAV from speaker AE in TIDIGITS) is shown in Fig.~\ref{fig:audio:events}.

\begin{figure}[!ht]
    \centering
    \includegraphics[width=\linewidth]{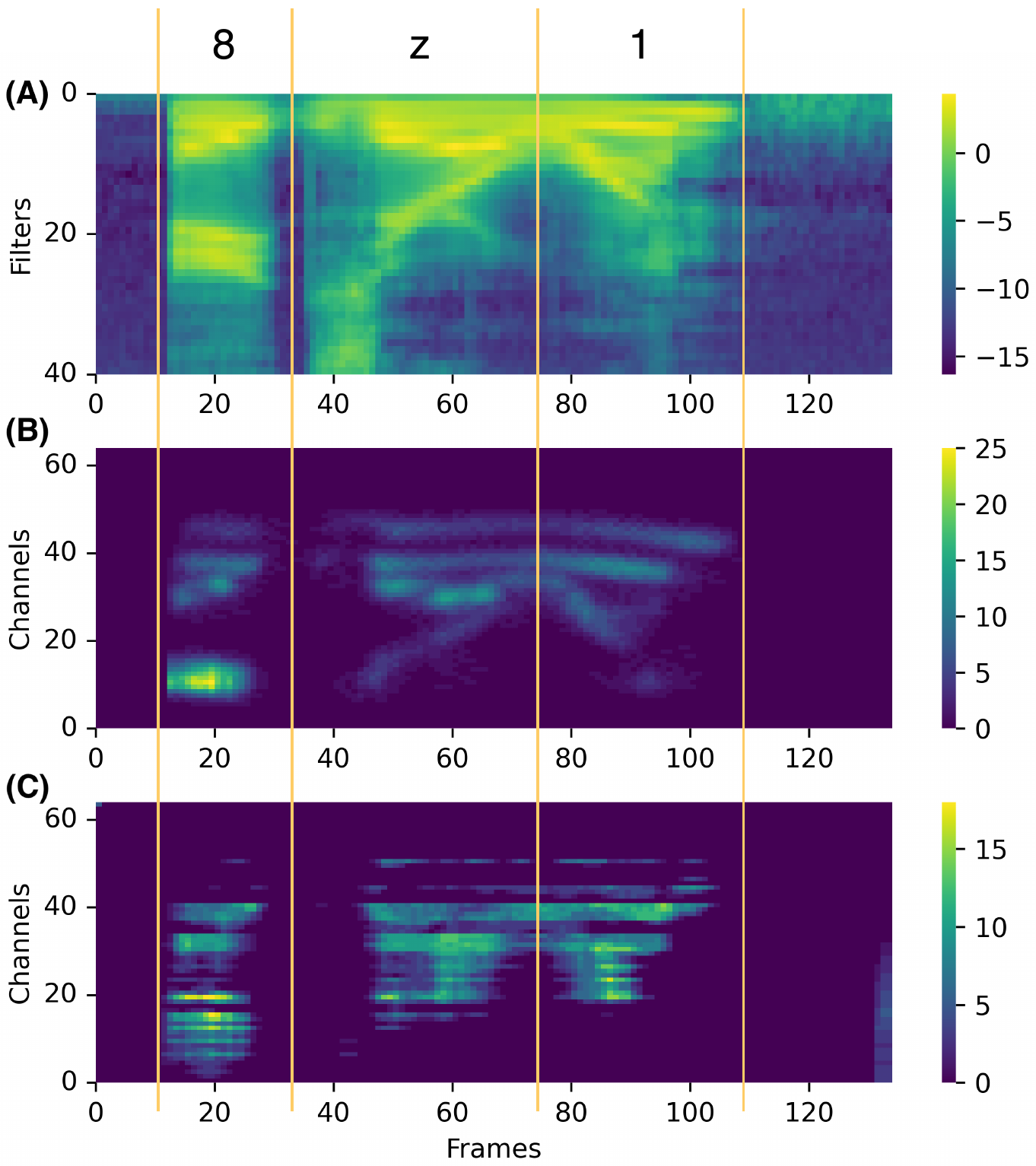}
    \caption{Features extracted from the audio sample ``8z1''. Log-Mel spectrogram features \textbf{(A)}. TBSC features  generated from simulated spikes \textbf{(B)} and DAS responses \textbf{(C)} using a configuration 25w/10s.}
    \label{fig:audio:features}
\end{figure}

\subsection{Feature Extraction} \label{subsec:audio:feature}

Features are extracted from both the audio and spike samples.
For audio samples, we computed the Log-Mel spectrogram features using a 25\,ms window, 10\,ms stride, and 40 filter bank channels as shown in Fig.~\ref{fig:audio:features}~(A).
For the spike recordings, we 
used the time-binned spike count (TBSC) method~\cite{event:feature:Anumula:2018} to convert spikes into event features.
Assume that the spike stream consists of $N$ spikes $\{e_{i}=(t_{i}, f_{i})\}_{i=1}^{N}$ where timestamps $t_{i}$ are in ascending order.
The TBSC feature for a time window $\Delta W$ is then defined as $\mathbf{s}_{j}^{(f)} = \text{card}(\{e_{i}|t_{i}\in j\text{-th window}, f_{i}=f\})$ where $\mathbf{s}_{j}^{(f)}$ represents TBSC feature value at $j$-th window for channel frequency $f$; $\text{card}(\cdot)$ is the cardinality of selected events. $\Delta W$ is 10\,ms or 25\,ms and the stride is 10\,ms for the TBSC features used in the study.
Examples of TBSC features with $\Delta W$= 25\,ms are shown in Fig.~\ref{fig:audio:features}~(B) and (C).
For convenience, feature configurations are coded as window/stride, \emph{e.g.}, 25w/10s means 25\,ms window and 10\,ms stride.

\subsection{Network Architecture}

The RNN used in this work consists of two Gated Recurrent Unit (GRU)~\cite{gru:Cho:2014} layers with 256 units each followed by a fully connected layer of 200 units with a LeakyReLU activation~\cite{leaky:relu:Mass:2013}, and a final classification layer of 12 output units. The network has a total of 677\,K parameters.

\subsection{Datasets}\label{subsec:dataset}

We used the men and women speaker recordings in the TIDIGITS dataset~\cite{tidigit:Leonard:1993} to match the spike recordings in the N-TIDIGITS18 dataset~\cite{event:feature:Anumula:2018}.
N-TIDIGITS18 consists of 8,623 training samples and 8,700 testing samples. Each sample consists of a single digit or multiple spoken digits. In total, there are 11 digits (``o'', ``zero'' and ``1'' to ``9'').

For T-NGA training, the N-TIDIGITS18 recordings and the corresponding audio samples were aligned temporally using the Dynamic Time Warping (DTW) algorithm~\cite{dtw:Giorgino:2009} before feature extraction. The spike recordings generated by the software spiking cochlea model are already aligned with the corresponding audio samples. 

\begin{table*}[ht]
    \centering
    \caption{Comparison of T-NGA with supervised training on an ASR task. Log-Mel spectrogram features are generated with a 25w/10s configuration. TBSC features are generated with 25w/10s or 10w/10s configurations. T-NGA results are highlighted in blue.}
    \label{tab:events:results}
    \begin{tabular}{lcccc}
        \hline
        \multicolumn{1}{c}{\multirow{2}{*}{\textbf{Model}}} & \multicolumn{1}{c}{\textbf{Training}} & \multicolumn{1}{c}{\multirow{2}{*}{\textbf{Features}}} & \textbf{Feature} & \multicolumn{1}{c}{\multirow{2}{*}{\textbf{WER (\%)}}}  \\
        & \multicolumn{1}{c}{\textbf{Method}} & & \textbf{Configuration} & \\
        \hline
        \multicolumn{5}{l}{\textbf{(A) on TIDIGITS dataset}} \\
        \hline
        \texttt{PT-25} & Supervised & Log-Mel & 25w/10s & 0.80$\pm$0.15 \\
        \hline
        \multicolumn{5}{l}{\textbf{(B) ideal software cochlea model on TIDIGITS dataset}} \\
        \hline
        \texttt{B-25} & Supervised & SW-TBSC\textsuperscript{\dag} & 25w/10s & 2.20$\pm$0.36\\
        \texttt{B-10} & Supervised & SW-TBSC & 10w/10s & 1.70$\pm$0.17\\
        \hline
        \rowcolor{myblue}\texttt{BGN-25} & T-NGA & Log-Mel+SW-TBSC & 25w/10s & 1.96$\pm$0.05\\
        \rowcolor{myblue}\texttt{BGN-10} & T-NGA & Log-Mel+SW-TBSC & 10w/10s & 1.70$\pm$0.04\\
        \hline
        \multicolumn{5}{l}{\textbf{(C) on N-TIDIGITS18 dataset}} \\
        \hline
        \texttt{S-25} & Supervised & TBSC & 25w/10s & 10.18$\pm$0.39\\
        \texttt{S-10} & Supervised & TBSC & 10w/10s & 9.63$\pm$0.20 \\
        Ref.~\cite{event:feature:Anumula:2018} & Supervised & TBSC & n/a & 13.9\% \\
        \hline
        \rowcolor{myblue}\texttt{GN-25} & T-NGA & Log-Mel+TBSC & 25w/10s & 15.01$\pm$0.82\\
        \rowcolor{myblue}\texttt{GN-10} & T-NGA & Log-Mel+TBSC & 10w/10s & 14.85$\pm$0.47\\
        \hline
        \hline
        \multicolumn{5}{l}{\dag SW: Software cochlea model; TBSC: Time-binned spike count features.} \\
        \hline
    \end{tabular}
\end{table*}

\section{Results} \label{sec:results}

We conducted experiments on a speech recognition task using the datasets described in Section~\ref{subsec:dataset}.
All networks in this section were trained for
50 epochs using the Adam optimizer~\cite{adam:Kingma:2014}.
For the supervised training experiments, we used a learning rate of 3e-4 and the Connectionist Temporal Classification (CTC) loss function~\cite{Graves:2006:CTC}. For T-NGA, we used a learning rate of 1e-3 and the T-NGA loss function in Eq.~\ref{eqn:tnga:loss}. Reported results show the mean and standard deviation of the word error rate (WER) for 5 runs of each experiment.

\subsection{T-NGA experiments}\label{sec:asr:results}

We first trained a network \texttt{PT-25} on TIDIGITS using spectrogram features as shown in Table~\ref{tab:events:results}(A).
The \texttt{PT-25} model was subsequently used as the pretrained model for T-NGA training to obtain grafted networks (GNs). For comparison, we also trained supervised networks (SNs) through standard supervised training on spiking cochlea datasets using event features.

Table~\ref{tab:events:results}(B) shows the results of using the event features created from ideal software spiking cochlea recordings.
The GNs \texttt{BGN-25}, \texttt{BGN-10} reached similar or better WERs than the SNs \texttt{B-25}, \texttt{B-10}.
These result show that T-NGA training could be advantageous as it does not require a labeled event dataset and still produces a network that shows equivalent accuracy.

Results on N-TIDIGITS18 presented in Table~\ref{tab:events:results}(C) show that the SNs \texttt{S-25}, \texttt{S-10} achieved $\sim$10\% WER while the GNs \texttt{GN-25}, \texttt{GN-10} reached $\sim$15\% WER. 
The higher WER of the GNs compared to the corresponding SNs is because the hardware spiking cochlea includes circuit non-idealites that affect the distribution of the output cochlea spikes, and the temporal mismatch between the audio samples and event sample still exists after DTW alignment. In Section~\ref{sec:results:analysis}, we show that the WER of GN increased when training on spike recordings generated by the same software cochlea model which includes circuit non-idealities.

Table~\ref{tab:events:results} also shows that both GNs and SNs perform better when using a 10w/10s event feature configuration, \emph{i.e.}, 10w/10s characterizes the spike recordings better than 25w/10s.
T-NGA training for \texttt{BGN-10} and \texttt{GN-10} models used 25w/10s Log-Mel spectrogram features and 10w/10s TBSC features. The results show that T-NGA can maximize the similarity between temporally aligned grafted states and pretrained states successfully while the feature configurations for the two modalities are different.

\subsection{Analysis} \label{sec:results:analysis}
We present two additional experiments to better understand the results obtained from T-NGA training and the resulting GN.

We first investigated quantitatively whether the circuit non-idealities of a spiking cochlea affect the performance of GN and SN. We generated another event dataset using the same software spiking cochlea model that includes circuit non-idealities from the the neuron spiking threshold, and filter quality factor variation. As a result, the WER of the GN was $\sim$3\% higher than the SN as shown in Table~\ref{tab:non:ideal:sw:cochlea}. This result reveals that these non-idealities in the hardware spiking cochlea contributed to the gap in WER between GNs and SNs as observed in Table~\ref{tab:events:results}(C).

\begin{table}[ht]
    \centering
    \caption{T-NGA and supervised training results using a software spiking cochlear model that includes circuit non-idealities. Feature configuration is 25w/10s.}
    \label{tab:non:ideal:sw:cochlea}
    \begin{tabular}{ccc}
        \hline
        \multicolumn{1}{c}{\textbf{Training Method}} & \multicolumn{1}{c}{\textbf{Features}} & \multicolumn{1}{c}{\textbf{WER (\%)}}  \\
        \hline
        Supervised & SW-TBSC & 5.31$\pm$1.07\\
        T-NGA & Log-Mel+SW-TBSC & 8.30$\pm$0.18 \\
        \hline
    \end{tabular}
\end{table}

\begin{figure}[ht]
    \centering
    \includegraphics[width=\linewidth]{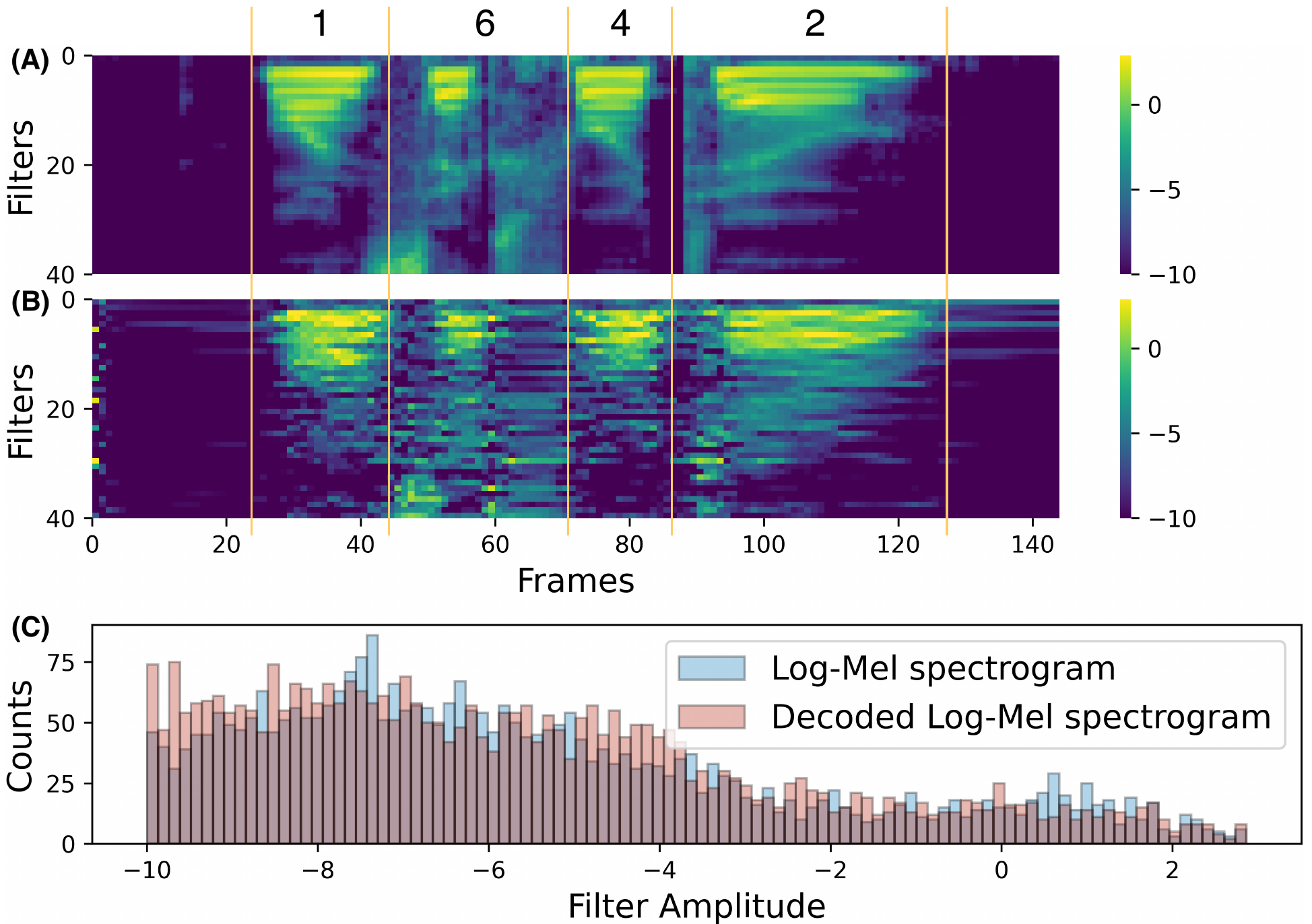}
    \caption{Log-Mel spectrogram features \textbf{(A)} computed from the sample "1-6-4-2". The decoded Log-Mel spectrogram features \textbf{(B)} are qualitatively similar to the features in \textbf{(A)}. The histogram \textbf{(C)} shows the distribution of filter amplitudes for the two feature sets.}
    \label{fig:decoded}
\end{figure}

We also explored the encoded content in the grafted states of GN by decoding a given set of grafted states through the pretrained front end by following~\cite{nga:Hu:2020}. The outcome of this decoding procedure is a decoded Log-Mel spectrogram that represents the given grafted states.
The optimization of this decoding procedure was carried out for 5,000 iterations using the Adam optimizer and a learning rate of 1e-2.
To suppress the noise in the decoded Log-Mel spectrogram, we clipped the amplitude at -10. Fig.~\ref{fig:decoded} shows a Log-Mel spectrogram that was decoded from the grafted states that were trained on a N-TIDIGITS18 sample. 
We found that the trained GN front end can generate the grafted features which often correctly represents the input audio signal.
The decoded spectrogram shown in Fig.~\ref{fig:decoded}~(B) is visually similar to the original Log-Mel spectrogram shown in Fig.~\ref{fig:decoded}~(A).  Fig.~\ref{fig:decoded}~(C) illustrates that the distributions of the amplitudes for the original and decoded spectrograms match well.

\section{Conclusion} \label{sec:conclusion}

We propose a self-supervised method called Temporal Network Grafting Algorithm (T-NGA) which exploits a RNN trained on audio spectrogram features to process event features through a newly trained grafted front end.
This algorithm mitigates the need for labeled spike recordings and only requires temporally aligned audio and spike samples.

We demonstrate T-NGA on a speech recognition task using TIDIGITS and N-TIDIGITS18 datasets; 
and show that the accuracy of the GN is equivalent to that of a network trained from scratch on the labeled software spiking cochlea recordings. When validated on real DAS recordings, the accuracy of the GN was only $\sim$5\% lower
than the accuracy of the SN despite the silicon circuit non-idealities reflected in the DAS recordings. The T-NGA works well even when the feature configuration of the Log-Mel spectrogram is different from that of the event features as shown in Section~\ref{sec:results}. Through self-supervised training, GN produces grafted states that are similar to the pretrained states as shown by the decoded spectrogram features.

Future work includes testing on a DAS dataset, which already includes the synchronized audio and spike recordings, and testing on datasets used for edge audio tasks such as keyword spotting.

\vfill\pagebreak

\bibliographystyle{IEEEbib}
\bibliography{refs}

\end{document}